\documentclass[article,pre,
twocolumn,
eqsecnum,
amsmath,amssymb,
amsmath,amssymb, aps,
]{revtex4}
\usepackage[dvipsnames]{xcolor} 
\usepackage{epsfig}
\usepackage{amsfonts}
\usepackage{amsmath}
\usepackage{wasysym}
\usepackage{amssymb}
\usepackage{bm} 
\usepackage[framemethod=PStricks]{mdframed}
\usepackage{lipsum}
\usepackage{float}

\usepackage{pdflscape}
\usepackage{graphicx}
\usepackage{ulem}

\newcommand{\be}{\begin{equation}}
\newcommand{\ee}{\end{equation}} 
\newcommand{\bea}{\begin{eqnarray}} 
\newcommand{\eea}{\end{eqnarray}}
\newcommand{\bfr}{{\bf{r}}}

\usepackage[utf8]{inputenc}
\usepackage{bbm} 				
\usepackage{epstopdf}				
\usepackage{verbatim}    			
\usepackage{sidecap}                
\usepackage{indentfirst}
\usepackage[colorinlistoftodos]{todonotes}

\begin{document}

\title{Multifractality Breaking from Bounded Random Measures}

\author{L. Moriconi}
\affiliation{ Instituto de F\'\i sica, Universidade Federal do Rio de Janeiro, \\
C.P. 68528, CEP: 21945-970, Rio de Janeiro, RJ, Brazil}


\begin{abstract}
Multifractal systems usually have singularity spectra defined on
bounded sets of Hölder exponents. As a consequence, their associated multifractal 
scaling exponents are expected to depend linearly upon statistical moment 
orders at high enough orders -- a phenomenon referred to as the 
{\it{linearization effect}}. Motivated by general ideas taken from 
models of turbulent intermittency and focusing on the case of 
two-dimensional systems, we investigate the issue within the framework of Gaussian 
multiplicative chaos. As verified by means of Monte Carlo simulations, it turns 
out that the linearization effect can be accounted for by Liouville-like random 
measures defined in terms of upper-bounded scalar fields. The coarse-grained 
statistical properties of Gaussian multiplicative chaos are furthermore found 
to be preserved in the linear regime of the scaling exponents. As a related 
application, we look at the problem of turbulent circulation statistics, and 
obtain a remarkably accurate evaluation of circulation statistical moments, 
recently determined with the help of massive numerical simulations.
\end{abstract}


\maketitle


\section{Introduction}

Soon after the realization that strange attractors should be
characterized by a set of generalized dimensions rather than
a single fractal dimension \cite{hentschel-procaccia,grassberger,schertzer-lovejoy},
related concepts were further developed and applied to the problem of homogeneous and 
isotropic turbulence \cite{benzi_etal,frisch-parisi,meneveau-sreenivasan1,
chhabra_etal,meneveau-sreenivasan2}. 
In the latter context, domains of the fluid velocity field which have
prescribed singular Hölder exponents have been conjectured to be fractal.
This is the essential content of the multifractal approach to turbulence,
which allows one to recover, within inertial range scales, the anomalous scaling 
properties of the turbulent energy cascade \cite{O62,K62,anselmet_etal,frisch}. 

The multifractal mindset has since then crossed the borders of its 
fluid dynamical birth place and is by now a valuable tool for the 
investigation of problems in fields as diverse as seismology, meteorology, 
ecology, condensed matter physics, dynamical systems, 
etc. \cite{sism_met,mat_cond,sis_din, eco,signals}.
In the particular case of turbulence, its worth emphasizing
that multifractality has been noted to be closely related to 
the Onsager's long-standing conjecture on flow singularities 
\cite{onsager,eyink_sreenivasan,eyink} and to the phenomenon 
of spontaneous stochasticity, a subject of growing interest, 
as far as it leads to a breaking of the deterministic paradigm of 
classical mechanics, in a sense which is even stronger than the one 
usually implied by chaotic behavior 
\cite{bernard_etal,chaves_etal, eyink_drivas, thalabard_etal,eyink_bandak}.

Multifractal modeling, however, is expected to be broken by extreme events: 
at enough high orders, statistical moments of the physical observables of
interest are found to depend linearly upon its moment orders, at variance with the
typical non-linear profiles predicted for multifractal systems \cite{molchan1,molchan2}.
While such a {\it{linearization effect}} can be actually explained 
in a natural way within the multifractal formalism \cite{ossiander_etal,Lahermes_etal1,
abry_etal, lashermes_etal2,muzy_etal,bacry_etal, angeletti_etal}, its account 
from alternative perspectives on multifractality has been puzzingly and,
as a consequence, a point of concern in applications. 
We have in mind, more specifically, the connection betweeen mutifractality 
and the theory of Gaussian multiplicative chaos  (GMC) \cite{kahane,rhodes_vargas}, 
which has been a fruitful tool in the development of finance \cite{duchon-robert}, 
turbulence \cite{chevillard_etal1,pereira_etal,chevillard_etal2,apol-mori}
and even quantum gravity models \cite{barral_etal}. Our aim, in this work, is to 
address the linearization effect in the framework of GMC 
and to illustrate, as a meaningful case study, an application of the proposed solution 
to the problem of turbulent circulation statistics \cite{Iyer_etal, migdal, apol_etal}.

This paper is organized as follows. In the next section, we clarify
the problem we are interested to study, recalling some 
relevant technical details of the multifractal formalism and the theory of 
GMC, with specific attention to 
the case of two-dimensional modeling. 
In Sec. III, heuristic arguments are introduced, based on phenomenological 
descriptions of turbulent cascades \cite{benzi_etal,O62,K62}, which motivate 
us to put forward, as a proposal, the necessary ingredients for the realization 
of the linearization effect along the lines of GMC.
Our conjectures are fully confirmed in Sec. IV by means of Monte Carlo 
simulations. We then carry out, in Sec. V, an application 
of the freshly derived results to the problem of circulation fluctuations 
in turbulence, obtaining excellent comparisons with evaluations obtained 
from previous numerical experiments \cite{Iyer_etal}. Finally, in Sec. VI, 
we summarize our findings and point out directions of further research.


\section{Problem Setup}

To set the stage for the issues we aim to address in this paper, let us consider 
the example of a $d$-dimensional positive-definite multifractal scalar field 
$\psi (x)$, described by some translation invariant probability measure, in
such a way that the statistical moments of the renormalized fields
\be
\psi_a(x)  \equiv \frac{1}{a^d} \int_{{\cal{D}}_a} d^d x \psi (x) \ , \  \label{psia}
\ee
where ${\cal{D}}_a$ is the spatial domain $|x' - x| \leq a$, 
behave as
\be
\mathbb{E}[(\psi_a(x))^q ] \sim a^{\tau_q} \label{psi_scaling} \ . \
\ee
The scaling exponents $\tau_q$ can be derived within the multifractal
language as follows \cite{frisch-parisi}. Let $D(h)$, referred to as the 
{\it{singularity spectrum}}, be the fractal dimension of the set of points 
which have Hölder exponent $h$ in the ensemble realizations of $\psi(x)$. 
Regarding $h$ as a random variable to be sorted when an arbitrary region 
of size $a$ is probed, the probability to find the local scaling 
behavior $\psi_a(x) \sim a^{h'}$ where $h' \in [h,h+dh]$ is, thus, $\rho(h) dh \sim a^{d -D(h)} dh$. 
We get, from these assumptions, that
\be
\mathbb{E}[(\psi_a(x))^q ] \sim \int dh \rho(h) a^{qh} \sim 
\int dh a^{hq+d-D(h)} \ . \ \label{multifractal_psia}
\ee
At small enough scales, the dependence of the above expectation values 
upon $a$ can be estimated with the help of the saddle-point method, which 
leads to (\ref{psi_scaling}), with
\be
\tau_q = {\hbox{inf}}_h [hq + d -D(h) ] \ . \ \label{tau_inf}
\ee
The scaling exponent $\tau_q$, therefore, is nothing but the
Legendre transform of the fractal codimension $d-D(h)$.

General arguments \cite{frisch} tell us that $\tau_q$ is a concave 
function of the moment order $q$. For the sake of 
clarity, we adopt here the convention that multifractality refers to the
case of scaling exponents which are strictly concave, that is, 
$d^2\tau_q/dq^2 < 0$. The singularity spectrum $D(h)$ 
is, accordingly, a strictly concave function of $h$.

As already alluded in the introductory section, multifractality is expected to be broken
at high enough moment orders. More concretely, this stands for the fact that for $q \geq q_c$, 
where $q_c$ is a model-dependent critical moment order, $\tau_q$ becomes a linear function of $q$ 
\cite{molchan1,molchan2,ossiander_etal,Lahermes_etal1,abry_etal,lashermes_etal2,
muzy_etal,bacry_etal,angeletti_etal}. An essential explanation of the linearization effect, 
apprehended from the aforementioned works, is that the domain 
of the singularity spectrum function is usually bounded from below by a limiting Hölder exponent $h_\ast$.
Therefore, taking into account that as $q$ grows, the value of $h$ which minimizes the 
RHS of (\ref{tau_inf}) gets smaller, it turns out that at some critical moment order $q_c$ 
the minimizer in (\ref{tau_inf}) saturates to $h_\ast$, leading to the monofractal relation
\be
\tau_q = h_\ast q + d -D(h_\ast)  \label{tau_infb} \ , \ 
\ee
for $q \geq q_c$.

The singularity spectrum can be well approximated in very many instances by a parabolic function of $h$ over a broad range of Hölder exponents, so that fluctuations of $\psi(x)$ can be effectively described by lognormal probability distribution functions. In equivalent words, the scaling exponents $\tau_q$ are given, in this approximation, by quadratic functions of $q$. In this connection, one notes that the combined existence of pointwise lognormal distributions for $\psi(x)$ and the scaling behavior of the coarse-grained variables (\ref{psi_scaling}) can be reproduced with the help of the Liouville measures as defined in the theory of GMC \cite{rhodes_vargas}. 

Centering our attention on two-dimensional modeling, the GMC approximation means, in practice, that $\psi(x)$ may be expressed as the Liouville measure density
\be
\psi(x) = \psi_0 \exp \left \{ \gamma \phi(x) -\frac{\gamma^2}{2} \mathbb{E}[\phi^2] \right \}  \ , \  \label{psi_gmc}
\ee
where $\psi_0 > 0$ and $\gamma$ are arbitrary parameters, and 
$\phi(x)$ is a free scalar field \cite{zinn-justin} with fluctuations
governed by the functional probability measure
\be
d \mu [\phi] = D[\phi] \exp \{ -S[\phi] \} \ , \ \label{prob_measure}
\ee
where
\be
S[\phi] = \frac{1}{2} \int_{{\cal{D}}_L} d^2 x (\partial_i \phi)^2   \ . \ \label{action}
\ee
Periodic boundary conditions are assumed for the scalar field $\phi(x)$ in the domain ${\cal{D}}_L$, which is furthermore discretized in a lattice of lattice parameter $\eta$ (a necessary technical detail for the ultraviolet regularization of the free field Green's functions). 
Taking $\eta \ll a \ll L$, it follows from (\ref{psia}) and (\ref{psi_gmc}-\ref{action}) that (\ref{psi_scaling}) 
is satisfied, that is,
\be
\mathbb{E}[(\psi_a(x))^q ] = c_q \psi_0^q \left ( \frac{a}{L} \right )^{\tau_q}  \ , \ \label{scale_a}
\ee
where $c_q$ is an unimportant dimensionless constant (for our
purposes) and
\be
\tau_q = \frac{\gamma^2}{4 \pi}q(1-q) \ . \ \label{tauq}
\ee
The bare field $\psi(x)$ can be identified with the ultraviolet regularized field $\psi_\eta(x)$.
A scaling relation similar to (\ref{scale_a}),
\be
\mathbb{E}[(\psi(x))^q ] = \left ( \frac{\eta}{L} \right )^{\tau_q}  \ , \  \label{scale_eta}
\ee
can then be derived from (\ref{psi_gmc}-\ref{action}) as well.

We wonder, thus, if it is possible to implement modifications in the GMC computational 
scheme based on Eqs. (\ref{psi_gmc}-\ref{action}) so as to get a crossover of the 
scaling exponents $\tau_q$, as $q$ grows, from (\ref{tauq}) to (\ref{tau_infb}), 
while still having power laws like (\ref{scale_a}) and (\ref{scale_eta}). In the next 
section, we propose a solution to this problem, relying on heuristic arguments inspired 
on well-known phenomenological models of turbulent intermittency.

The apparent methodological restriction represented by the use of turbulence phenomenology should 
not be a matter of concern at all,  since multifractal phenomena and techniques are usually traded 
without much difficulty among models of completely different nature.

\section{Bounded Cascades}

We briefly outline, in the following subsections A and B, two phenomenological views on the 
turbulent cascade, which when placed {\it{vis a vis}} the theory of GMC and the multifractal 
formalism, give relevant hints on how to establish the linearization effect in the framework 
of the GMC, a task addressed in subsection C.

\subsection{The Obukhov-Kolmogorov lognormal model of turbulent intermittency}

If $\psi(x)$ is used to model the turbulent dissipation field in homogeneous and isotropic
turbulence, commonly denoted by $\epsilon(x)$, relations (\ref{psia}) and (\ref{psi_scaling}) 
yield a precise formulation of the Kolmogorov refined similarity hypothesis, a central point 
in the Obukhov-Kolmogorov (OK62) modeling of turbulent intermittency \cite{O62,K62}. 

In the OK62 phenomenology, multiplicative cascade fluctuations of the energy transfer 
rates per unit mass and unit time, $\epsilon_a$ and $\epsilon_b$, across two different 
length scales $a$ and $b$, respectively, are related as
\be
\epsilon_a = \epsilon_b W_1W_2...W_n \ . \ \label{cascade}
\ee
The $W's$ are lognormally i.i.d. random variables, with unit mean,
and $n = \log_2 (b/a) $, taken to be a positive integer, 
gives the number of modeled steps in the turbulent energy cascade between the scales 
$a$ and $b$. They are assumed to lay within the inertial range scales, that is, 
$\eta \leq a < b \leq L$, where $L$ and $\eta$ define the integral and dissipative 
length scales, respectively, of the turbulent flow. 
Eq. (\ref{cascade}) is to be understood in the probabilistic sense as an equality 
in law for $\epsilon_a$ and $\epsilon_b$.

Considering $b=L$, that is, the scale where energy is injected into the flow with 
non-fluctuating energy transfer rate $\epsilon_L$, then it is a straightforward exercise 
to show, from (\ref{cascade}), that 
\be
\mathbb{E}[\epsilon_a^q] \sim \epsilon_L^q \left( \frac{a}{L} \right )^{\tau_q}   \label{eps-tauq}
\ee
holds for $\eta \leq a \leq L$, in the same fashion as (\ref{scale_a}) and 
(\ref{scale_eta}), where $\tau_q$ is given as in (\ref{tauq}), with
\be
\gamma^2 = 2 \pi \log_2 \mathbb{E}[W^2] \ . \
\ee
As a relevant note for future use, we introduce the Gaussian random variable $X_p$, through 
\be
W_p \equiv \exp( \gamma X_p) \ . \  \label{Wp}
\ee
The OK62 cascade argument (\ref{cascade}) can in this way be recalled to
suggest, taking a look at (\ref{psi_gmc}), that pointwise fluctuations of 
$\psi(x)$ can be derived from $\psi \sim \exp(\gamma \phi)$, where
\be
\phi = \sum_{p=1}^{\log_2(L/\eta)} X_p \ . \  \label{phiXp}
\ee

\subsection{The random $\beta$-model of turbulent intermittency}

An alternative OK62-like cascade picture of the energy transfer rate fluctuations 
across scales, as synthetized in Eq. (\ref{cascade}), can be put forward in order 
to render it closer to contemporary multifractal ideas and in compliance with 
general physical principles like energy conservation.

In the random $\beta$-model \cite{benzi_etal}, an arbitrary energy-containing eddy defined 
at length scale $a$ produces, during its lifetime, a random number $M_a \leq 2^d$ 
of descendent eddies (in $d$ dimensions), all of them defined at length scale $a/2$. 
Energy conservation implies that the power supplied by the mother-eddy 
to its descendents has to be same as the total power supplied by 
the latter ones to their further descendents and, as a consequence,
$M_a \epsilon_{a/2} (a/2)^d =  \epsilon_a a^d$, that is
\be
\epsilon_{a/2} = \beta_a^{-1} \epsilon_a \ , \ 
\ee
where $\beta_a = M_a / 2^d$ is the fraction of volume that the whole group of
descendent eddies (the ``sibling-eddies") occupy with the respect to the volume 
of their mother-eddy.

Assuming that generation after generation the $\beta's$ are completely independent
and randomly distributed according to the same probability density function $f(\beta)$,
Eq. (\ref{cascade}) still holds for the energy transfer rates of each individual eddy, with
(subindices suppressed)
\be
W = \beta^{-1} \ . \ \label{Wbeta}
\ee
We have, therefore,
\be
\epsilon_a = \left [ \prod_{i=1}^n \beta_i^{-1} \right ] \epsilon_L \ , \ 
\ee
where $n = \log_2 (L/a)$.  Energy transfer rates have, now, statistical moments
\bea
&&\mathbb{E}[\epsilon_a^q] \sim  \epsilon_L^q 
\left [ \int_0^1 d \beta f(\beta) \beta^{ 1 - q} \right ]^n \nonumber \\
&&= \epsilon_L^q \mathbb{E}[\beta^{ 1 - q} ]^n
\sim \epsilon_L^q \left( \frac{a}{L} \right )^{\tau_q} \ , \ \label{random-beta}
\eea
where
\be
\tau_q = - \log_2 \mathbb{E}[\beta^{ 1 - q} ] \ . \  \label{tauq-randombeta}
\ee
Note that the expectation value (\ref{random-beta})
takes into account the fact that in the random $\beta$-model 
eddies are not space-filling structures.

The particular modeling case where $\beta$ is fixed to some arbitrary value 
$\beta_0$ \cite{frisch_beta}, associated to the probability distribution function
\be
f(\beta) = \delta(\beta - \beta_0) \ , \
\ee
gives, in view of (\ref{tauq-randombeta}), the linear scaling exponents
\be
\tau_q = (q-1) \log_2 \beta_0 \ . \  \label{tauq-randombeta2}
\ee
Furthermore, still considering the situation of fixed $\beta$, we infer that 
a mother-eddy at the integral scale $L$ (the ``mother of all mothers") is the source, 
along the turbulent cascade, of a number
\be
N_a \sim (2^d \beta_0)^{\log_2(L/a)} \sim \left ( \frac{L}{a} \right )^{d + \log_2 \beta_0} \label{fractal_scal}
\ee
of descendent eddies at length scale $a$. The scaling law (\ref{fractal_scal}) 
indicates that the fractal dimension of the energy-containing eddies is, here,
\be
d_F = d + \log_2 \beta_0 \ . \ \label{fracdim}
\ee

\subsection{The linearization effect in the theory of GMC}

As discussed in Sec. II, the linearization 
effect takes place when statistical moments get dominated by fluctuations 
associated to the most singular set of configurations, which are the ones
which have the minimum available Hölder exponent, denoted in Eq. (\ref{tau_infb}) 
by $h_\ast$. Due to the concavity properties of the singularity spectrum, 
we expect the fractal dimension of the most singular set, $D(h_\ast)$, 
to be the smallest allowed one (for the evaluation of positive order moments).

We also note that Eq. (\ref{fracdim}) can be used to establish a mapping 
between values of $\beta$, from the side of the random $\beta$-model, to 
the fractal dimensions encompassed by the singularity spectrum $D(h)$, 
from the side of the multifractal formalism. In the language
of the random $\beta$-model, the linearization effect follows from the
existence of a minimum value of $\beta$, say $\beta_\ast$, obtained from
\be
D(h_\ast) = d + \log_2 \beta_\ast \  . \
\ee
Taking (\ref{Wbeta}) into account, we conclude that the cascade factors
$W's$ are, under these conditions, upper bounded random variables, viz., 
$W \leq 1/ \beta_\ast$. Correspondingly, we see,
from the context of the OK62 phenomenology,
that bounded $W's$ should be related to bounded scalar fields 
$\phi(x)$ in the GMC setup, as indicated by (\ref{Wp}) and 
(\ref{phiXp}).

Relying upon the above heuristic considerations, we are, now, 
ready to propose a modified version of the two-dimensional GMC, 
as given by Eqs. (\ref{psi_gmc}-\ref{action}), 
in order to accommodate in its formal structure the linearization 
effect. To do so, we actually keep the definition of the functional 
probability measure (\ref{prob_measure}), but
\vspace{0.2cm}

\noindent (i) replace the Liouville measure (\ref{psi_gmc}) by the more general expression
\be
\psi(x) = \frac{\psi_0}{\mathbb{E}[\tilde \psi(x)]} \tilde \psi(x) \ , \ \label{psi_gmcb}
\ee
where 
\be
\tilde \psi(x) = \exp [ \gamma \phi(x) ] \ ; \ \label{psi_gmcc}
\ee
\vspace{0.2cm}

\noindent (ii) replace the Euclidean action, Eq. (\ref{action}), by
\be
S[\phi] = \int d^2 x \left [  \frac{1}{2}  (\partial_i \phi)^2 + V(\phi) \right ] \ , \ \label{actionb}
\ee
where
\begin{equation}
V(\phi) =
    \begin{cases}
      0 \ , \ \text{if $\phi < \phi_0 $} \ , \ \\
    V_0 \ , \ \text{if $\phi \geq \phi_0$} \ , \ \\
    \end{cases}    \label{V}
\end{equation}
with $V_0 \rightarrow   \infty$ and 
\be
\phi_0 = C \ln(L/\eta) \ , \ \label{phi0}
\ee
where $C$ is an adjustable positive constant (observe that
(\ref{phi0}) follows from (\ref{phiXp}) by taking
$X_p = C \ln 2$). 
\vspace{0.2cm}

In short words, we have just postulated that 
the Liouville measure (\ref{psi_gmcb}) gets upper bounded due to the 
existence of a scalar field threshold $\phi_0$, and that it fluctuates 
as usually determined by the free field action (\ref{action}), if 
$\phi(x) < \phi_0$ in an arbitrary neighborhood of $x$.

An analytical treatment of the modified GMC scenario, as defined by 
Eqs. (\ref{psi_gmcb}-{\ref{phi0}), 
is challenging. However, it is possible to proceed along with
Monte Carlo numerical validations, as detailed next.

\section{Monte Carlo Simulations}

We have performed Monte Carlo simulations to study the fluctuations of
the non-normalized Liouville measure (\ref{psi_gmcc}), with $\gamma=1$, 
using (\ref{actionb}-\ref{phi0}), for the {\it{pure GMC}} ($\phi_0 = \infty$) 
and {\it{modified GMC}} ($\phi_0 < \infty$) cases. 

Statistical ensembles with configurations of $\tilde \psi (x)$
have been produced for systems of three different sizes: 
$L/\eta = 30, 50,$ and $100$,
through the application of the standard Metropolis algorithm 
\cite{binder-heermann}. An educated guess 
for the value of $C$ in (\ref{phi0}) gives
\be
\phi_0 > \sqrt{\mathbb{E}_0[\phi^2]} \ , \  \label{phi0C}
\ee
where $\mathbb{E}_0[\cdot]$ stands for expectation values taken in 
the pure GMC scheme. The rationale for (\ref{phi0C}) is that 
at low enough orders, statistical moments of $\tilde \psi (x)$ are 
expected to be approximately described by quadratic scaling exponents 
like the ones of the pure GMC case, since the scalar field $\phi(x)$
will very rarely fluctuate beyond the standard deviation range, 
$\sqrt{\mathbb{E}_0[\phi^2]}$. On the other hand, as the moment order 
grows, larger fluctuations of $\phi(x)$ come into play, reaching more 
frequently the upper bound $\phi_0$, thus opening the way to the onset 
of the linearization effect. A direct computation yields
\be
\mathbb{E}_0[\phi^2] = \frac{1}{2 \pi} \ln \left ( \frac{L}{\eta} \right ) \ . \ 
\ee
Taking, $C \equiv 2/\ln(30)$, one can then easily check that the inequality 
(\ref{phi0C}) holds in fact for all the studied system sizes.

Each Monte Carlo run consisted of $10^7$ iterations, sampled at every other 
10 steps, which evolved from the initial state $\phi(x) =0$. 
The field derivatives in the action ({\ref{actionb}) were 
evaluated by means of central differences. Monte Carlo variations 
of $\phi(x)$ (defined at lattice sites) were given by independent 
pseudorandom numbers uniformly distributed in the interval $[-1,1]$. 

Statistical moments of the bare and the coarse-gained non-renormalized
Liouville measures, $\tilde \psi(x)$ and 
\be
\tilde \psi_a(x)  \equiv \frac{1}{a^d} \int_{{\cal{D}}_a} d^d x \tilde \psi (x) \ , \  \label{tilde_psia}
\ee
respectively, are reported in Figs. 1 and 2. As
evidenced from Fig. 1, the linearization effect is well reproduced in the 
modified GMC framework for the moment order range $5 \leq q \leq 11$
($q=11$ is the largest analysed order). Fig. 1 also shows the excellent 
collapse of data for the investigated systems, which supports the 
finite-size dependent definition of the upper bound (\ref{phi0}).

\begin{figure}[ht]
\hspace{0.0cm} \includegraphics[width=0.45\textwidth]{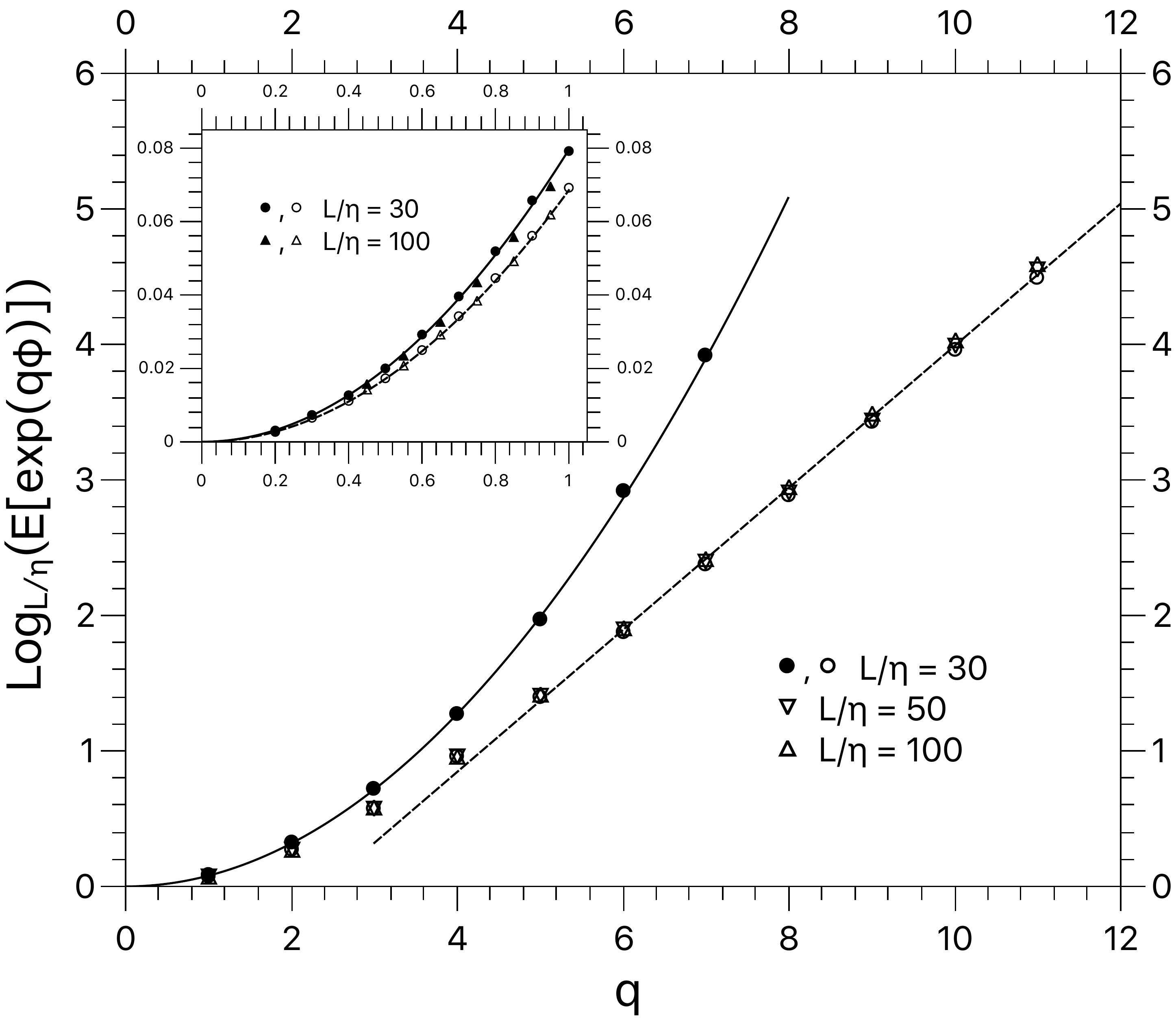}
\vspace{0.0cm}
\caption{Solid lines/symbols and open symbols refer, respectively, to the 
pure and modifed GMC cases. Symbols yield scaling exponents obtained from 
Monte Carlo simulations. The solid line (either in the main plot or
the inset) is the predicted parabolic scaling exponent 
$q^2/4 \pi$, for the moments of (\ref{psi_gmcc})
in the pure GMC case, as it can be derived from Eq. (\ref{tauq}),
with $\gamma =1$. Dashed lines (in the main plot and inset) are 
linear and parabolic 
fits.} 
\label{}
\vspace{0.4cm}

\hspace{0.0cm} \includegraphics[width=0.45\textwidth]{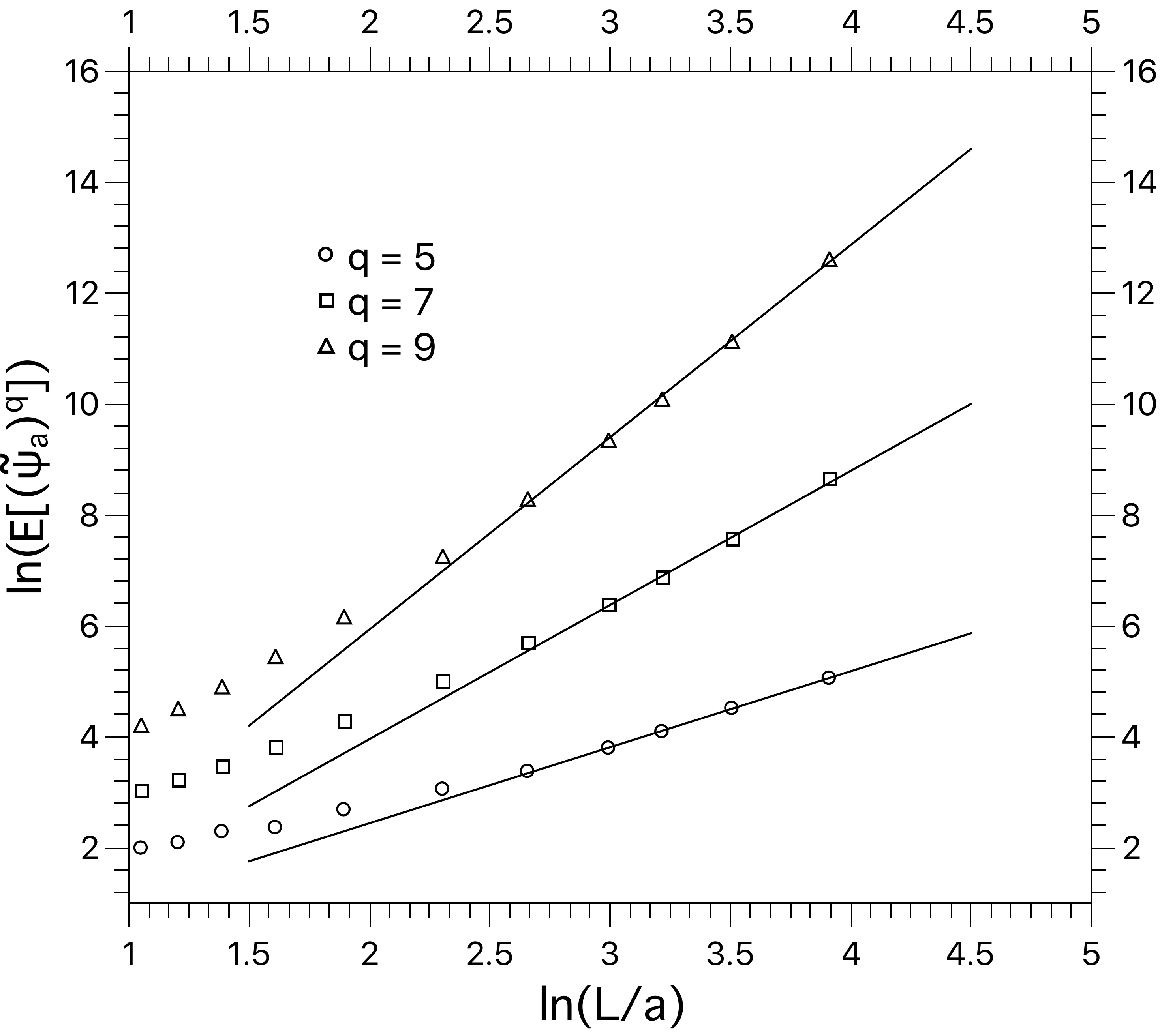}
\vspace{0.0cm}
\caption{Scaling regimes for the moments of the coarse-grained 
non-normalized Liouville measures $\tilde \psi_a(x)$, as evaluated 
from Monte Carlo simulations (symbols) performed on a system of size
$L/\eta = 100$. The solid straight lines have slopes which are 
the scaling exponents obtained from the linear fit shown in Fig. 1, 
for the moment orders $q=5,7,$ and $9$.} 
\label{}
\end{figure}

The Monte Carlo results depicted in Fig. 2 indicate that 
$\tilde \psi_a(x)$ scales with the same scaling exponent as 
$\tilde \psi(x) \equiv \psi_\eta(x)$ at small length scales 
($a/L < 0.1$), even for moment orders where the linearization 
effect is observed. 

The linearization effect for the coarse-grained bounded 
Liouville measures is a remarkable phenomenon, which has
an immediate impact in turbulence modeling, since it
bridges the linearization effect for scaling quantities like the 
velocity structure functions to the linearization effect for the
turbulent dissipation field, if one assumes, of course,
that the Kolmogorov refined similarity hypothesis is still valid. 
We examine, in the following, this interesting phenomenological point 
in connection with a recently discussed model for the turbulent 
fluctuations of the circulation variable \cite{apol_etal}.

\section{Turbulent Circulation Statistics}

The relevance of the circulation variable \cite{acheson} as a multiscale ``mathematical probe" of turbulent vortical 
structures, pointed for the first time some 25 years ago \cite{migdal2}, has recently found renewed interest with the advent of
high performance computing and improved data storage capability \cite{Iyer_etal}. Novel modeling ideas have 
been put forward \cite{Iyer_etal, migdal, apol_etal}, including possible connections between the statistics 
of circulation in classical and quantum turbulent flows \cite{muller_etal}.

Let us center our attention on the particular definition of circulation as
\be
\Gamma_R \equiv \int_\mathcal{C} d^2 \bfr \,  \omega(\bfr)  \ , \ \label{circ} 
\ee
where  $\mathcal{C}$ is a disk of radius $R$ and $\omega(\bfr)$ is the component of vorticity which is normal 
(with arbitrary orientation) to the plane that contains $\mathcal{C}$. The scaling form for the circulation moments,
\be
\mathbb{E}[|\Gamma_R|^q] \sim R^{\lambda_q} \ , \ \label{circ_scaling}
\ee
is observed to hold for the inertial range of scales $\eta \ll R \ll L$ \cite{Iyer_etal}. 
We are here mainly interested to model the scaling exponents $\lambda_q$ in (\ref{circ_scaling}).
The Kolmogorov phenomenological description of turbulence (K41) \cite{frisch} yields 
$\lambda_q = 4q/3$, which has been noted to be a very good approximation only for 
$q \leq 4$ \cite{Iyer_etal}.

Tracing back circulation fluctuations to the presence of vortex tubes, it was proposed, in 
Ref. \cite{apol_etal}, that the vorticity field in (\ref{circ}) can be effectively represented, 
for the purpose of evaluating the statistical moments (\ref{circ_scaling}), as
\be
\omega(\bfr) \sim \xi_R \tilde  \omega(\bfr) \ , \  \label{omega}
\ee
where
\be
\xi_R \equiv \frac{1}{\pi R^2} \int_\mathcal{C} d^2 \bfr  \, \sqrt{ \epsilon(\bfr)} \label{xsi}
\ee
is a functional of the dissipation field $\epsilon(\bfr)$, modeled as a Liouville measure density,
and $\tilde  \omega(\bfr)$ is an independent Gaussian random field, with vanishing mean
and correlator 
\be
\mathbb{E}[\tilde  \omega(\bfr) \tilde  \omega(\bfr')] \sim |\bfr - \bfr'|^{- \alpha} \ . \ \label{omegacorrelator}
\ee
The scaling exponent $\alpha$ in (\ref{omegacorrelator}) can be determined, as we will see in a 
moment, from the imposition of general phenomenological constraints \cite{comment}.

\begin{figure}[ht]
\hspace{0.0cm} \includegraphics[width=0.45\textwidth]{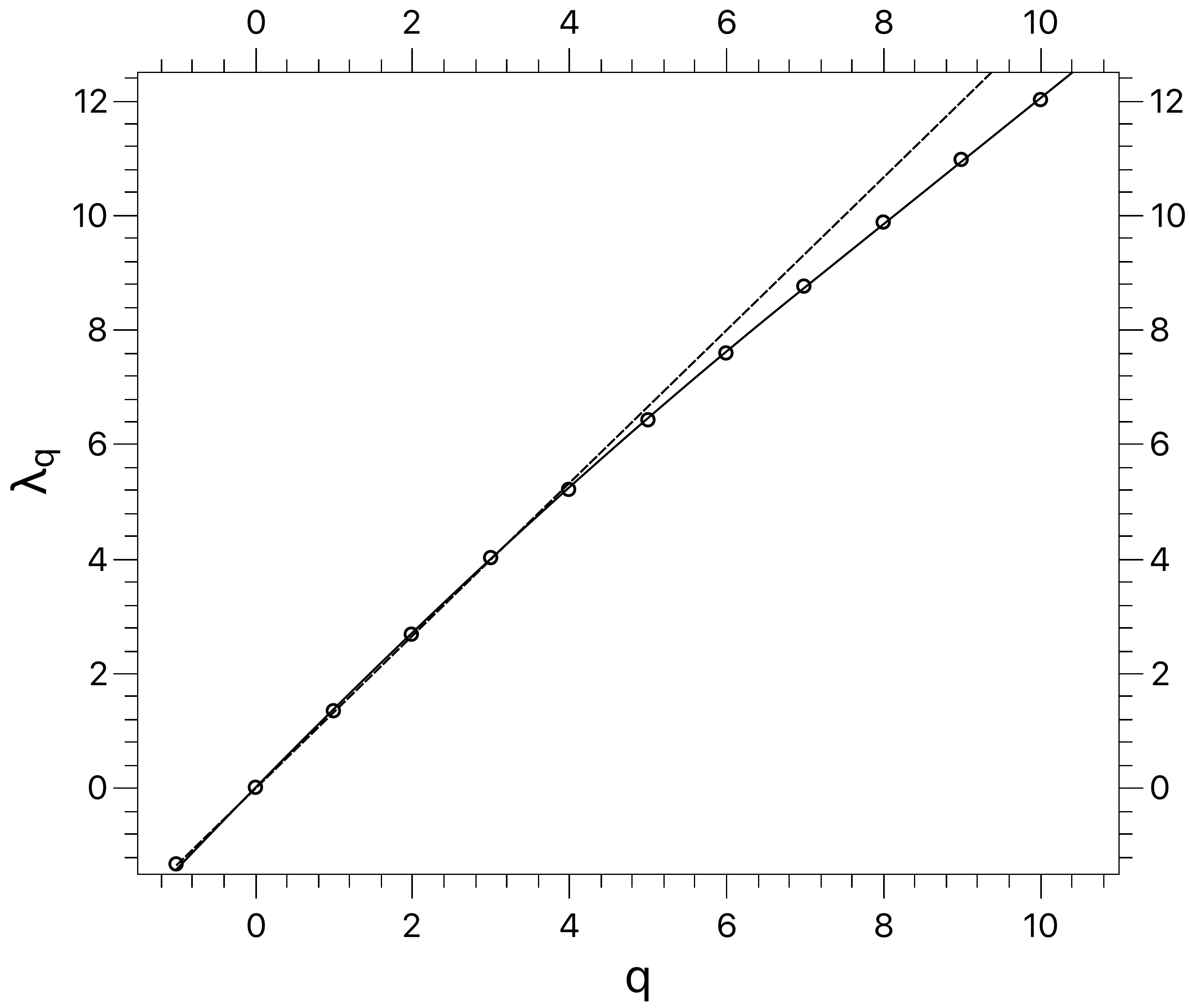}
\vspace{0.0cm}
\caption{Scaling exponents for the circulation moments. 
Symbols give the values obtained through direct numerical simulations \cite{Iyer_etal}. The dashed line
is the K41 linear profile, $\lambda_q = 4q/3$, while the solid line is the prediction of the present model,
as given in Eq. (\ref{lambdaqb}).} 
\label{}
\end{figure}

Since powers of Liouville measures are Liouville measures as well, as it can be clearly seen 
from the definition (\ref{psi_gmcc}), we are able to obtain (\ref{circ_scaling}) by putting 
together (\ref{circ}), (\ref{omega}), and the coarse-grained Liouville measure (\ref{xsi}), 
with
\be
\lambda_q = \tau_{q/2} + (4 - \alpha) \frac{q}{2} \ , \ \label{lambdaq}
\ee
where $\tau_q$ is the energy transfer rate exponent formally introduced in (\ref{eps-tauq}).
We determine, now, the crossover moment order $q_c$ that defines the onset of
the linearization effect. For $q < q_c$ the OK62 lognormal model gives \cite{O62,K62,frisch}
\be
\tau_{q/2} = \frac{\mu}{8} q (2-q) \ , \ \label{tauq2}
\ee
where $\mu = 0.17 \pm 0.01$ \cite{tang_etal}. The relation between $q_c$ and the Hölder
exponent minimizer $h_\ast$ (associated to singularities of the dissipation field) 
can be worked out without much difficulty; we get
\be
q_c = 1 - 2 \frac{h_\ast}{\mu} \ . \  \label{qchast}
\ee
Determinations of the singularity spectrum of the energy dissipation field 
from high Reynolds number experiments was accomplished in Ref. \cite{meneveau-sreenivasan1,chhabra_etal}. 
It turns out, from a careful analysis of the reported data, that 
$h_\ast \simeq -0.5$. This leads us, from (\ref{qchast}), 
to $q_c \simeq 6.88$.

It remains to discuss the yet undetermined exponent $\alpha$. Considering 
that there is no anomalous scaling for the third order velocity structure functions, as 
signalized in Kolmogorov's 4/5 law \cite{frisch}, we postulate that $\lambda_3 = 4$, exactly
as in K41 phenomenology \cite{comment2}. Using (\ref{lambdaq}) and (\ref{tauq2}) with $q=3$,
we obtain, thus, 
\be
\alpha = \frac{4}{3} - \frac{\mu}{4} \ . \ 
\ee
Collecting all the above pieces of information, we write down the circulation scaling exponent as
\begin{equation}
\lambda_q =
    \begin{cases}
    \bar  \lambda_q \equiv \frac{4}{3}q + \frac{\mu}{8} q (3-q)  \ , \ \text{if $q < q_c $} \ , \ \\
    \frac{1}{2} \left ( h_\ast + \frac{8}{3} + \frac{\mu}{4} \right ) (q-q_c) + \bar \lambda_{q_c} \ , \ \text{if $q \geq q_c$} \ . \
    \end{cases}    \label{lambdaqb}
\end{equation}

The comparison of the predicted values of $\lambda_q$ with the results of massive numerical simulations \cite{Iyer_etal} is 
excellent, as shown in Fig. 3.  The transition in behavior of the statistical moments of circulation as their moment orders 
are varied was actually observed for the first time in Ref. \cite{Iyer_etal}. We see, therefore, that it can be consistently 
explained as a manifestation of multifractality breaking, or, in other words, the linearization effect, within the modeling arena of GMC.

\section{Conclusions}

We have been able to address a variation of GMC, as described from relations (\ref{psi_gmcb}-\ref{phi0}), 
which gives room for the linearization effect, a phenomenon commonly observed in multifractal systems. The key technical point in the definition of the modified GMC setting is the introduction of upper-bounded Liouville measures. 

Our line of reasoning has been closely motivated by cascade models of turbulent intermittency and their connections with the mulifractal language and the theory of GMC. We validated the modified picture of GMC by means of straightforward Monte Carlo simulations and applied it to the problem of turbulent circulation statistics. We developed, in this way, accurate evaluations of the scaling exponents for the 
statistical moments of circulation, previously established only through extensive numerical simulations \cite{Iyer_etal}. 

Further work is in order. It would be very interesting to devise a mathematically rigorous analysis of the
linearization effect in the GMC, as discussed in this work, and to extend it to general space dimensions. Additional 
Monte Carlo simulations are also welcome to explore the validity range (as the field bound $\phi_0$, the system size, 
and moment orders are changed) of the modified GMC picture. The empirical (numerical) implementation of bounded Liouville measures to models based on the theory of GMC should not present any relevant technical or conceptual difficulty.
\vspace{0.2cm}

\leftline{\it{Acknowledgments}}
\vspace{0.2cm}

The author thanks G.B. Apolinário, R.M. Pereira, and V.J. Valadão for several enlightening discussions.
This work was partially supported by CNPq.



\begin{references}
\bibitem{comment} The value $\alpha = 4/3$ was actually taken in Ref. \cite{apol_etal} as a first approximation, 
based on K41 phenomenology. Here, we carry out a slighlty more general discussion avoiding to fix $\alpha$ 
from the outset.
\end{references}
\end{document}